\documentclass[fleqn,10pt]{wlscirep}
\usepackage[utf8]{inputenc}
\usepackage[T1]{fontenc}
\usepackage{bm}
\usepackage{siunitx}
\usepackage{lettrine}

\definecolor{dcol}{rgb}{0.7, 0.1, 0.1}

\newcommand{\aj}{Astron. J.}   
\newcommand{\apj}{Astrophys. J.}   
\newcommand{\apjl}{Astrophys. J. Lett.}   
\newcommand{\aap}{Astron. Astrophys.}   
\newcommand{\mnras}{Mon. Not. R. Astron. Soc.}   
\newcommand{\nat}{Nature} 
\newcommand{\nastro}{Nat. Astron.} 
\newcommand{\pasa}{Publ. Astron. Soc. Aust.}   
\newcommand{\pasp}{Publ. Astron. Soc. Pac.}   

\providecommand{\arcmin}{\hbox{$^\prime$}}
\providecommand{\arcsec}{\hbox{$^{\prime\prime}$}}
\providecommand{\degr}{\hbox{$^{\circ}$}}
\providecommand{\farcs}{\mbox{\ensuremath{.\!\!^{\prime\prime}}}}
\providecommand{\fs}{\hbox{\ensuremath{.\!\!^{\rm s}}}}
\newcommand{\dms}[3]{\ensuremath{{#1}\degr\,{#2}\arcmin\,{#3}\arcsec}}
\newcommand{\hms}[3]{\ensuremath{{#1}^{\rm h}\,{#2}^{\rm m}\,{#3}^{\rm s}}}
\newcommand{\dmsf}[4]{\ensuremath{{#1}\degr\,{#2}\arcmin\,{#3}\,{\farcs}{#4}}}
\newcommand{\hmsf}[4]{\ensuremath{{#1}^{\rm h}\,{#2}^{\rm m}\,{#3}\,{\fs}{#4}}}

\definecolor{orcid}{rgb}{0.06, 0.4, 1.0}
\newcommand{\orcid}[2]{\href{https://orcid.org/#2}{ \textcolor{orcid}{#1}}}

\title{A radio technosignature search towards Proxima Centauri resulting in a signal-of-interest}

\author[1,2]{\orcid{Shane Smith}{0000-0003-4101-8234}}
\author[3,2]{\orcid{Danny C. Price}{0000-0003-2783-1608}}
\author[4,2]{\orcid{Sofia Z. Sheikh}{0000-0001-7057-4999}}
\author[2]{\orcid{Daniel J. Czech}{0000-0002-8071-6011}}
\author[2,5]{\orcid{Steve Croft}{0000-0003-4823-129X}}
\author[6]{\orcid{David DeBoer}{0000-0003-3197-2294}}
\author[2]{\orcid{Vishal Gajjar}{0000-0002-8604-106X}}
\author[2,7]{\orcid{Howard Isaacson}{0000-0002-0531-1073}}
\author[2]{\orcid{Brian C. Lacki}{0000-0003-1515-4857}}
\author[2]{\orcid{Matt Lebofsky}{0000-0002-7042-7566}}
\author[6]{David H.E. MacMahon}
\author[2,5,8]{\orcid{Cherry Ng}{0000-0002-3616-5160}}
\author[9]{\orcid{Karen I. Perez}{0000-0002-6341-4548}}
\author[2,5,10]{\orcid{Andrew P.V. Siemion}{0000-0003-2828-7720}}
\author[11,2]{Claire Isabel Webb}
\author[12]{Jamie Drew}
\author[12]{S. Pete Worden}
\author[13, 14]{\orcid{Andrew Zic}{0000-0002-9583-2947}}

\affil[1]{Department of Physics, Hillsdale College, 33 E College St, Hillsdale, MI 49242, USA}
\affil[2]{Department of Astronomy, University of California Berkeley, Berkeley CA 947203, USA}
\affil[3]{International Centre for Radio Astronomy Research, Curtin University, Bentley WA 6102, Australia}
\affil[4]{Department of Astronomy \& Astrophysics, Pennsylvania State University, University Park, PA 16802, USA}
\affil[5]{SETI Institute, 89 Bernardo Ave, Suite 200 Mountain View, CA 94043, USA}
\affil[6]{Radio Astronomy Laboratory, University of California Berkeley, Berkeley, CA 94523, USA}
\affil[7]{University of Southern Queensland, Toowoomba, QLD 4350, Australia}
\affil[8]{Dunlap Institute for Astronomy \& Astrophysics, University of Toronto, 50 St.  George Street, Toronto, Ontario, M5S3H4, Canada}
\affil[9]{Department of Astronomy, Columbia University, 550 West 120th Street, New York, NY 10027, USA}
\affil[10]{Dept. of Physics \& Astronomy, University of Manchester, Oxford Road, M139PL, UK}
\affil[11]{Berggruen Institute, Los Angeles CA 90013, USA}
\affil[12]{Breakthrough Initiatives, NASA Research Park, Moffett Field, CA 94043, USA}
\affil[13]{Department of Physics and Astronomy, and Research Centre in Astronomy, Astrophysics and Astrophotonics, Macquarie University, NSW 2109, Australia}
\affil[14]{Australia Telescope National Facility, CSIRO, Space and Astronomy, PO Box 76, Epping, NSW 1710, Australia}

\affil[*]{e-mail: dancpr@berkeley.edu}

\begin{abstract}
The detection of life beyond Earth is an ongoing scientific endeavour, with profound implications. One approach, known as the search for extraterrestrial intelligence (SETI), seeks to find engineered signals (`technosignatures') that indicate the existence technologically-capable life beyond Earth. Here, we report on the detection of a narrowband signal-of-interest at $\sim$982\,MHz, recorded during observations toward Proxima Centauri with the Parkes Murriyang radio telescope. This signal, `BLC1', has characteristics broadly consistent with hypothesized technosignatures and is one of the most compelling candidates to date. Analysis of BLC1---which we ultimately attribute to being an unusual but locally-generated form of interference---is provided in a companion paper\cite{Sheikh:2021:BLC1}. Nevertheless, our observations of Proxima Centauri are the most sensitive search for radio technosignatures ever undertaken on a star target. 
\end{abstract}
\begin{document}

\flushbottom
\maketitle

\thispagestyle{empty}


\lettrine{T}{he}  discovery of the exoplanet Proxima Centauri b (Proxima~b) in orbit around Proxima Centauri\cite{2016Natur.536..437A} has sparked excitement over the prospect of a habitable exoplanet in the nearest reaches of the solar neighborhood. Several studies\cite{2016A&A...596A.111R,2016A&A...596A.112T, Alvarado-Gomez:2020} suggest that Proxima~b may be able to sustain an atmosphere favorable for life. However, because Proxima Centauri is an active M-dwarf flare star, doubt has been cast on the ability of Proxima~b, which is in a much tighter orbit than Earth is to the Sun, to retain an atmosphere amendable to the existence of biological life. A naked-eye visible superflare strong enough to kill any known organisms has been observed from Proxima Centauri\cite{2018ApJ...860L..30H}, although  life could still exist on the cold side of a tidally-locked planet. Coronal mass ejections from Proxima Centauri have also been observed\cite{Zic:2020,MacGregor:2021}, suggesting Proxima~b experiences significant ionizing radiation. Nevertheless, there remain compelling arguments that M-dwarf stars are viable hosts for life-bearing planets\cite{Tarter:2007}, and Proxima~b remains a compelling target for biosignature and technosignature searches.

Proxima~b is also a candidate for \textit{in situ} searches for extraterrestrial life by Breakthrough Starshot\cite{Parkin:2018}. Starshot seeks to launch a gram-sized, laser-light-sail propelled spacecraft at a relativistic velocity (0.2c) to the $\alpha$ Centauri system with Proxima~b as the primary target. If successful, the spacecraft will travel the $4.22$ light years to Proxima Centauri in 20 years and then transmit information of its target(s) back to Earth.

Despite being our nearest stellar neighbor, few technosignature searches have been conducted toward Proxima Centauri. In the 90's, two SETI programs were conducted in the Southern hemisphere toward nearby stars: the Project Phoenix search of 202 Solar-like stars\cite{Backus:1995, Tarter:1996}, and a search for technosignatures from 176 of the brightest stars\cite{Blair:1992}. Consequently, neither program selected Proxima Centauri---a faint M-dwarf---as a star target. Until a recent technosignature search of high-resolution optical spectra of Proxima Centauri\cite{Marcy:2021}, no technosignature searches had been conducted at optical wavelengths. This search of archival data from the HARPS spectrometer between 2004 and 2019 would have revealed laser emission from Proxima Centauri with $<$120\,kW power, and was motivated by the observations detailed here.

Here, we present a search for technosignatures from the direction of Proxima Centauri, using the CSIRO Parkes radio telescope (`Murriyang') as part of the Breakthrough Listen (BL) and the Breakthrough Initiatives search for life beyond Earth. BL---a sister initiative of Starshot---is a 10-year program to search for signs of intelligent life at radio and optical wavelengths \cite{2017AcAau.139...98W, 2017PASP..129e4501I}. BL has been conducting observations since 2016 and is undertaking the most rigorous and comprehensive observational SETI campaign to date \cite{Enriquez:2017, 2020AJ....159...86P, Sheikh:2020, Traas:2021, Gajjar:2021}. Proxima Centauri is part of the BL survey sample of 60 stars within 5\,pc \cite{2017PASP..129e4501I}, and was observed as part of a previous data release including 1327 nearby stars\cite{2020AJ....159...86P}. These observations were conducted using the Parkes 10-cm receiver, covering 2.60--3.45\,GHz. The observations presented here cover a larger bandwidth (0.7--4.0 GHz), with $6\times$ longer on-source dwell times (30 minutes). Our observations allow us to set the lowest equivalent isotropic radiated power (EIRP) detection limit for any stellar target.

We searched our observations toward Proxima Centauri for signs of technologically-advanced life, across the full frequency range of the receiver (0.7--4.0\,GHz). To search for narrowband technosignatures we exploit the fact that signals from any body with a non-zero radial acceleration relative to Earth, such as an exoplanet, solar system object, or spacecraft will exhibit a characteristic time-dependent drift in frequency (referred to as a drift rate) when detected by a receiver on Earth. We applied a search algorithm that detects narrow-band signals with doppler drift rates consistent with that expected from a transmitter located on the surface of Proxima~b (Fig.~1). Our search detected a total of 4,172,702 hits---i.e. narrow-band signals detected above a signal-to-noise (S/N) threshold---in all on-source observations of Proxima Centauri and reference off-source observations. Of these, 5,160 hits were present in multiple on-source pointings toward Proxima Centauri, but were not detected in reference (off-source) pointings toward calibrator sources; we refer to these as `events' (see Tab.\,1). 

\begin{figure*}
    \centering
    \includegraphics[width=0.6\columnwidth]{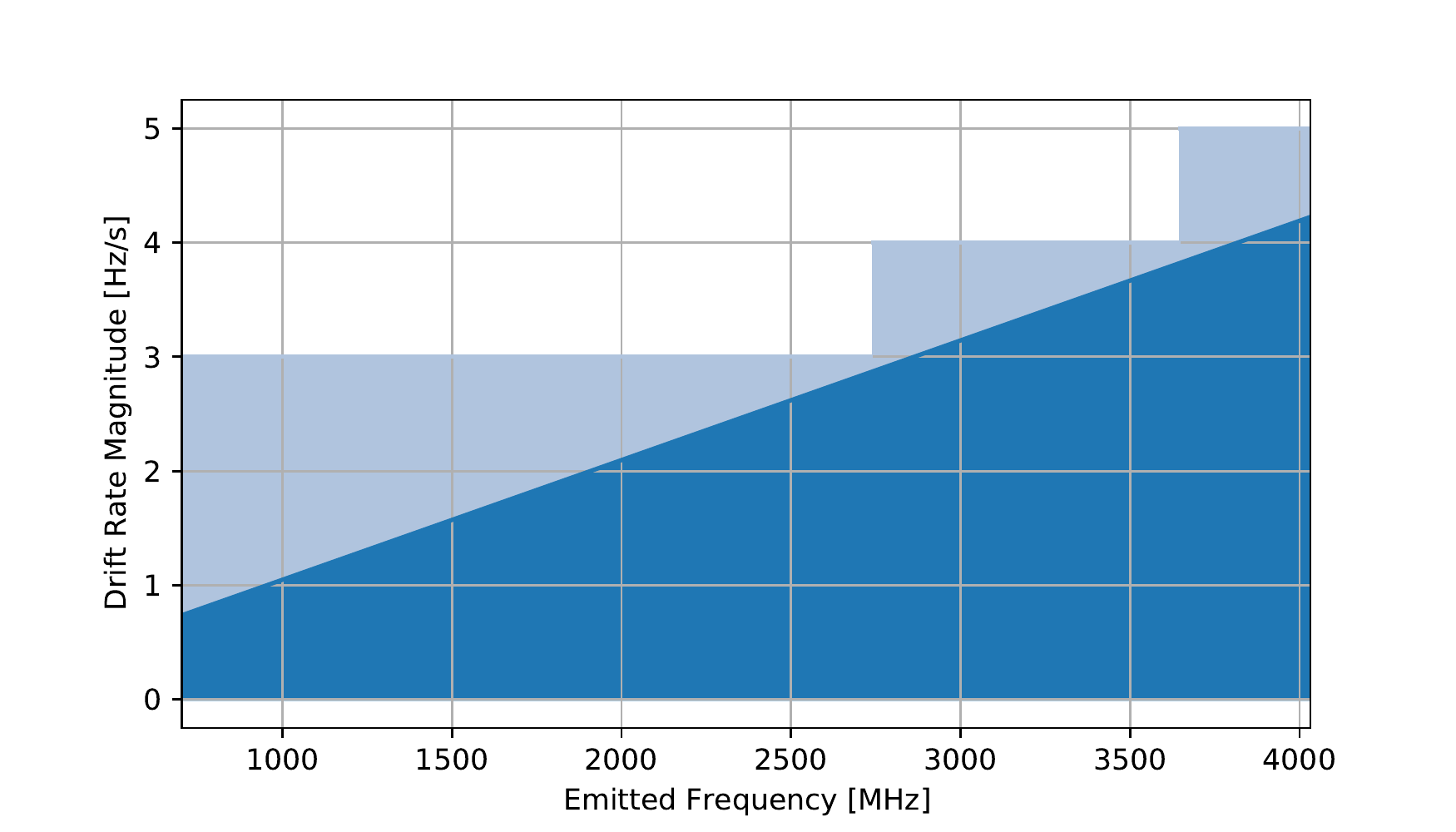}%
    \caption{Expected Doppler drift from a transmitter located on the surface of Proxima~b. Dark blue is the calculated magnitude of the Doppler drift and light blue is the magnitude of the search range.\label{fig:expected-doppler}}
\end{figure*}

The total hits by frequency are shown in Fig.~2. As expected, our detection pipeline finds the majority of hits (57\%) in ranges that have registered transmitters. A distribution of hits and events for our drift rate search range and S/N are shown in Fig.~3. Positive, negative, and zero drift rates correspond to 10\%, 15\%, and 75\% for the total hits respectively. The slight bias towards a negative drift rate is due to non-geosynchronous satellites\cite{2006JNav...59..293Z}. The majority of events occur below a S/N threshold of $10^3$ because faint signals are less likely to be detected in our shorter reference observations. Stronger signals are also generally associated with nearby ground-based transmitters that will appear in both on-source and off-source observations.

Out of the 5,160 events, only one event (Fig.~4) passed all rounds of filtering and visual inspection of dynamic spectra. The event does not lie within the frequency range of any known local radio-frequency interference (RFI), and has many characteristics consistent with a putative transmitter located in another stellar system. This event, which we refer to as a signal-of-interest, has been previously reported as `BLC1', short for `Breakthrough Listen Candidate 1'. We note that `signal-of-interest' is more appropriate than `candidate'\cite{Forgan:2019}, but for consistency we will adopt BLC1 throughout. 

BLC1 was detected at 982.002571\,MHz, with a drift rate of 0.038\,Hzs$^{-1}$. The signal-of-interest was detected over a $\sim$2.5\,hr period, and is only present in pointings toward Proxima Centauri. According to the US Federal Communications Commission (FCC) and the Australian Radiofrequency Spectrum Plan (ARSP), BLC1 lies within a frequency band reserved for aeronautical radionavigation; however, no transmitters that operate at the detected frequency of BLC1 are registered within 1000\,km of the observatory. Radionavigation stations are ground-based, so are less likely to be directionally sensitive. It is unlikely that an aircraft or satellite would be present in the direction of Proxima Centauri over the course of the signal-of-interest. BLC1 is analysed in further detail in a companion paper \cite{Sheikh:2021:BLC1}. As detailed in the companion paper, we ultimately conclude that BLC1 is a complex intermodulation product of multiple human-generated interfers: not a technosignature.

\begin{figure*}
    \centering
    \includegraphics[width=1.0\columnwidth]{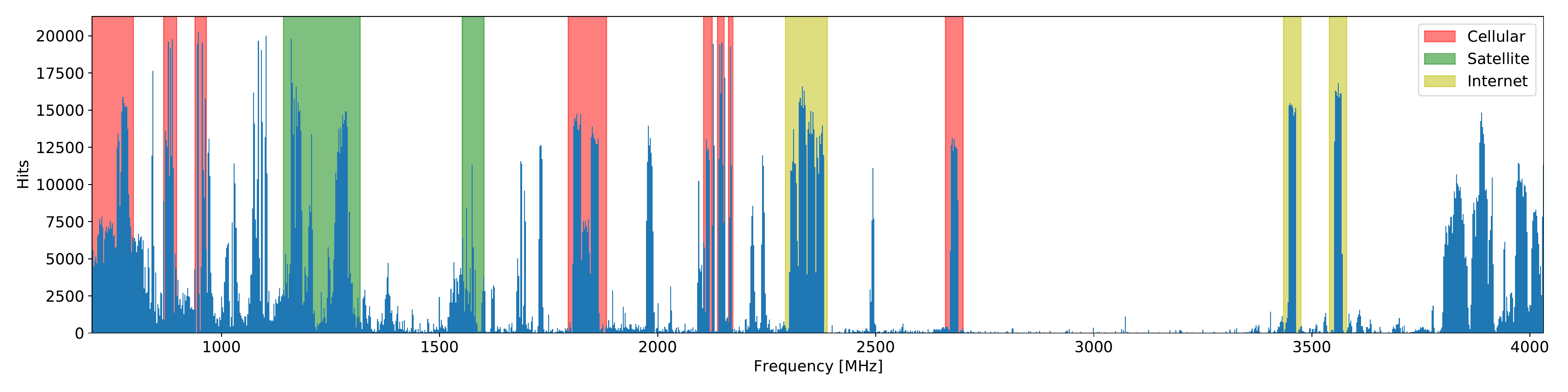}%
    \caption{A histogram of total hits as a function of frequency (narrowband signals detected above S/N threshold) for our observations toward Proxima Centauri. We used the \textsc{turboSETI} doppler search code to search for narrowband signals with a doppler drift across the 0.7--4.0\,GHz bandwidth of the Parkes UWL receiver. Registered cellular, satellite and broadband internet transmitters in the Parkes area overlaid. 
\label{fig:allhits}}
\end{figure*}

Given our non-detection of technosignatures, we place limits on the detection of narrow-band signals from Proxima Centauri by calculating the minimum detectable EIRP ($\mathrm{EIRP_{min}}$). The $\mathrm{EIRP_{min}}$ is given by 
\begin{equation}
	 \mathrm{EIRP_{min}} = 4\pi d^2F_{\mathrm{min}} \label{eq:EIRP}
\end{equation}
where d is the distance to the source (1.301\,pc for Proxima Centauri) and $F_{\mathrm{min}}$ is the minimum detectable flux in \si{W/m^{2}}. The equation for $F_{\rm{min}}$ depends on the minimum S/N ($\mathrm{S/N_{min}}$), the system temperature of the telescope ($T_{\mathrm{sys}}$), the effective collecting area of the telescope ($A_{\mathrm{eff}}$), the channel bandwidth (B), the number of polarizations ($n_{\mathrm{pol}}$), and the total observation time ($t_{\mathrm{obs}}$)\cite{2020AJ....159...86P,2017ApJ...849..104E}:
\begin{equation}
	 \mathrm{F_{min}} = \mathrm{S/N_{min}}\frac{2k_BT_{\mathrm{sys}}}{A_{\mathrm{eff}}}\sqrt{\frac{B}{n_{\mathrm{pol}}t_{\mathrm{obs}}}}.\label{eq:Fmin}
\end{equation}
We calculate $F_{\mathrm{min}} = 9.2$\,Jy\,Hz and $\mathrm{EIRP_{min}} = 1.9$\,\si{GW}. This $\mathrm{EIRP_{min}}$ is $3.6\times$ smaller than that previously reported $\mathrm{EIRP_{min}} = 6.2$\,GW for observations of Proxima Centauri\cite{2020AJ....159...86P}. The improved $\mathrm{EIRP_{min}}$ is due to the lower $T_{\mathrm{sys}}$ of the UWL receiver (22\,K), compared to the Parkes 10-cm receiver (35\,K), and our longer $t_{\mathrm{obs}}$, 5\,min versus 30\,min. Additionally, our $\mathrm{EIRP_{min}}$ is $1.6\times$ smaller than Green Bank Telescope L-band (1.1--1.9\,GHz) and S-band (1.7--2.6\,GHz) observations of the second closest star outside of the Alpha Centauri system, Barnard's star (GJ\,699). As such, our search of Proxima Centauri is decisively the most sensitive and comprehensive technosignature search done for a stellar target.

Based on previous SETI searches and analysis\cite{Wlodarczyk-Sroka:2020}, it is clear that putative narrowband transmitters are rare. As such, it is statistically probable that any signal-of-interest is a pathological case of human-generated interference. Extended and rigorous analysis of BLC1 was required to ascertain its progeny; this is presented in the companion paper\cite{Sheikh:2021:BLC1}, alongside a framework for verification of future signals-of-interest. 

Alone, this search---or more generally, any band-limited single-target search---cannot disprove the existence of a technologically advanced society on Proxima~b. While the UWL receiver has a wide bandwidth, we have still not covered the entire radio spectrum, nor optical, infrared, or X-ray bands. In addition to false positives, RFI could also confound detection of technosignatures at coincident frequencies. Proxima Centauri remains an interesting target for technosignature searches, and we encourage continued observations with other facilities and alternative approaches.

\section*{Methods}

For the observations of Proxima Centauri presented here, we used the Parkes Ultra-Wide bandwidth, Low-frequency receiver (UWL)\cite{2020PASA...37...12H}. The receiver has an effective system temperature ($T_{\rm{sys}}$) of 22\,K and system equivalent flux density (SEFD) of 28\,Jy across $\sim 60$\%\ of the band. The UWL receiver covers a 3.3\,GHz wide bandwidth from 0.704 to 4.032\,GHz. 

Observations were conducted over UT 2019 April 29 to UT 2019 May 4, as part of the P1018 project, `Wide-band radio monitoring of space weather on Proxima Centauri'.  In this project, Parkes observations were part of a multi-wavelength campaign in which Proxima Centauri was monitored for stellar flare activity \cite{Zic:2020}. Observations were conducted using the BL Parkes digital recorder \cite{Price:2018,Price:2021} (BLPDR) in parallel with the primary UWL digital signal processor, `Medusa'. For the purposes of a narrowband technosignature search, we deal solely on Stokes-I data from BLPDR. For the purposes of detecting flare activity, data from Medusa---which was configured to produce a full-Stokes data product at high time resolution ($128 \mu$s) but low frequency resolution (1\,MHz)---were also recorded; however, these data are not included in our analysis given their relative insensitivity to narrowband signals.

\begin{figure*}
    \centering
    \includegraphics[width=0.6\columnwidth]{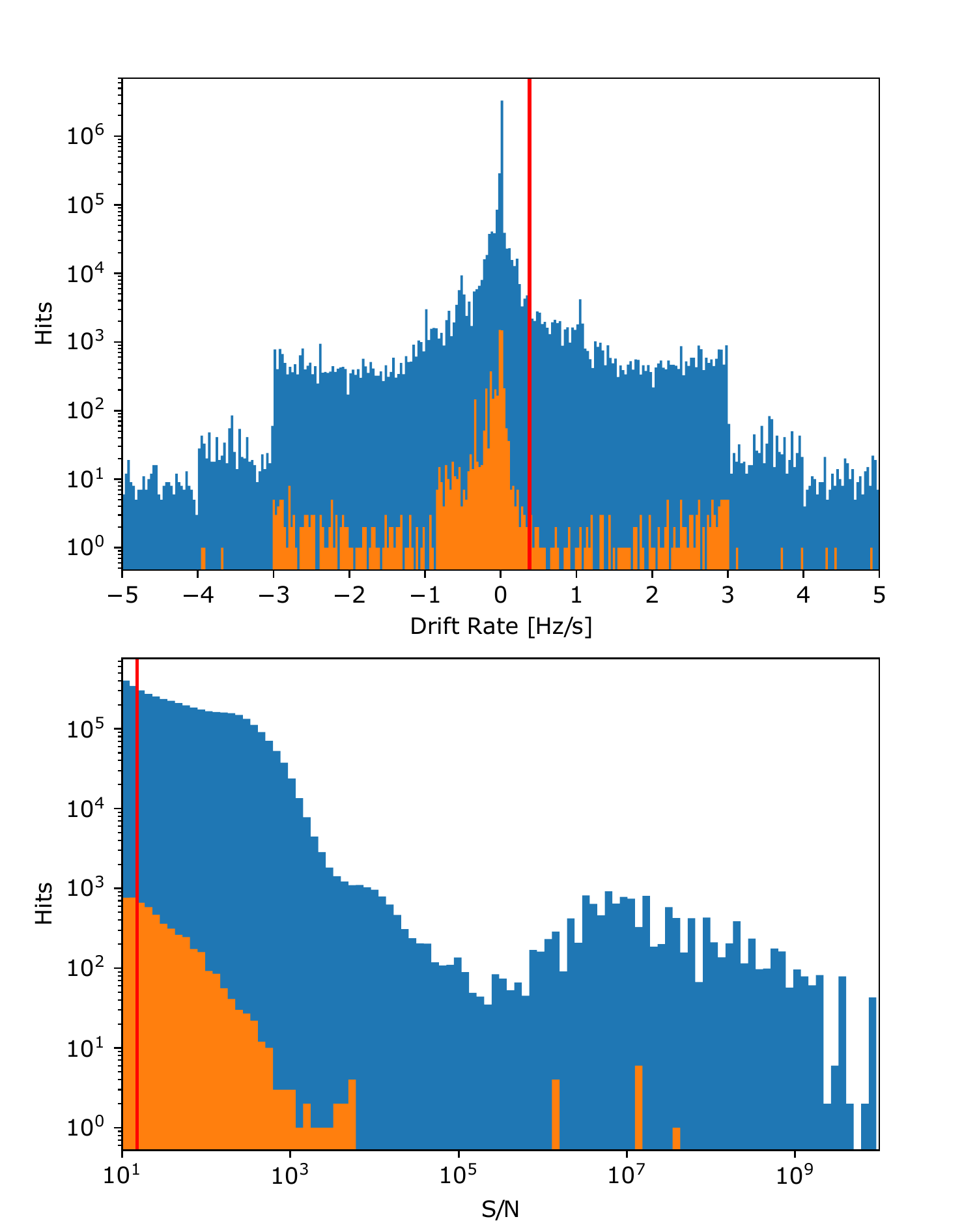}%
    \caption{Histograms with total hits and events for the search drift rates (top panel) and signal-to-noise (bottom panel). Blue bins are hits (signals detected by our data analysis pipeline) and orange bins are events (signals detected in all on-source observations, but not in off-source observations). \label{fig:snr-drate}}
\end{figure*}

\begin{figure*}
    \centering
    \includegraphics[width=1.0\columnwidth]{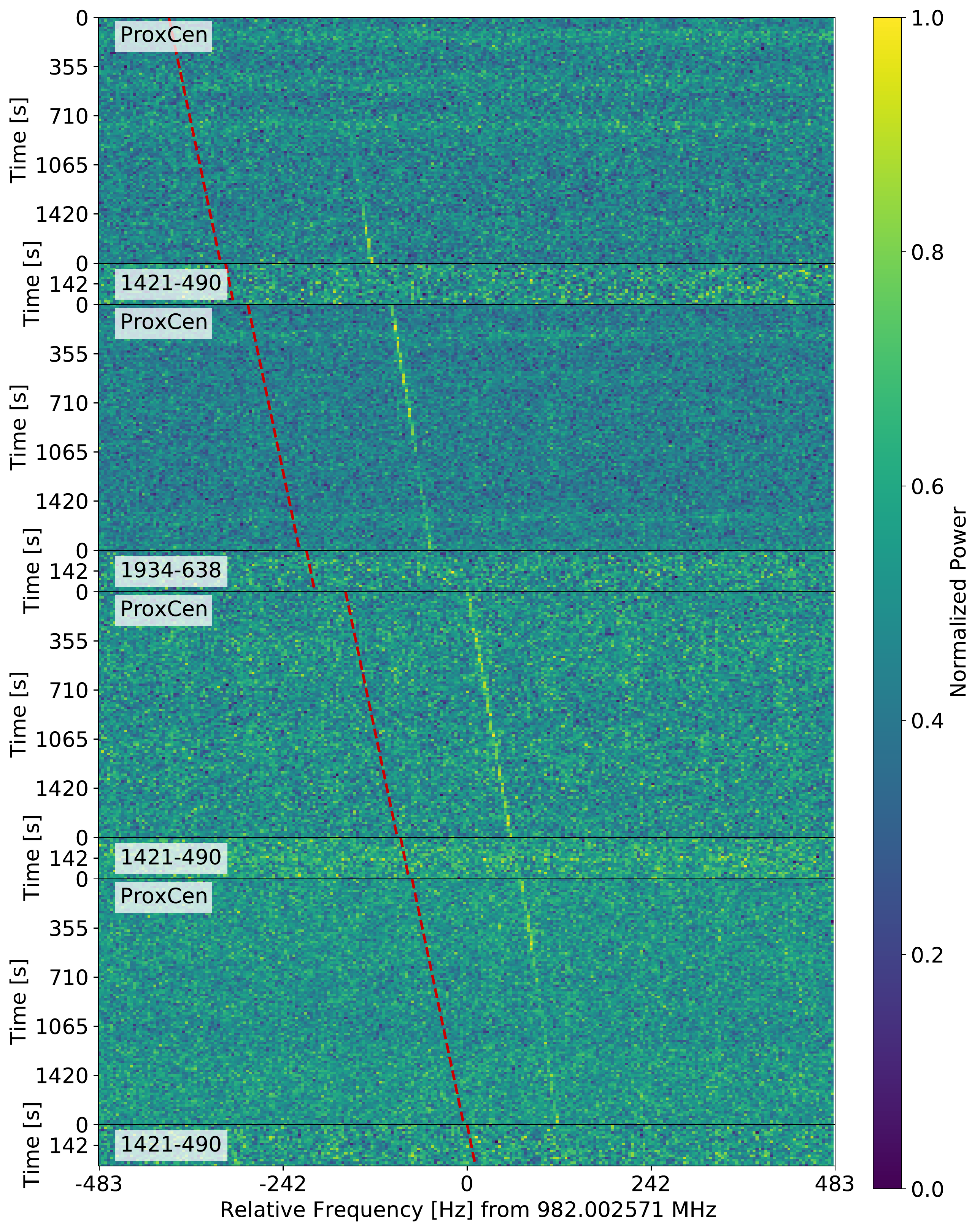}%
    \caption{The signal-of-interest, BLC1, from our search of Proxima Centauri. Here, we plot the dynamic spectrum around the signal-of-interest over an  8-pointing cadence of on-source and off-source observations. BLC1 passes our coincidence filters and persists for over 2 hours. The red dashed line, purposefully offset from the signal, shows the expected frequency based on the detected drift rate (0.038 Hz\,s$^{-1}$) and start frequency in the first panel. BLC1 is analyzed in detail in a companion paper\cite{Sheikh:2021:BLC1}. \label{fig:top-cand}}
\end{figure*}

\subsection*{Observation Strategy}
Over the period UT 2019 April 29 to UT 2019 May 4, we observed Proxima Centauri for a total of 26 hours and 9 minutes. The observations were conducted as a series of on-source, off-source pointings (a `cadence') to enable rejection of RFI, similar to previous BL observations\cite{2020AJ....159...86P, 2017ApJ...849..104E}. The on-source pointings (A) were toward Proxima Centauri (\hmsf{14}{29}{42}{95}, \dmsf{-61}{59}{53}{84}) while the off-source pointings (B) were primarily the calibrator source PKS\,1421-490 (\hms{14}{24}{32.24}, \dms{-49}{26}{50.26}) with the exception of two of the pointings, which used PKS\,1934-638 (\hmsf{19}{39}{25}{03}, \dmsf{-62}{57}{54}{34}), a well-characterized flux calibrator for Parkes.

Our observations differed from standard BL searches in two ways. First, our observation lengths were $\sim 30$-m on-source and $\sim 5$-m off-source; previous BL searches\cite{2017ApJ...849..104E,2020AJ....159...86P} employed a 5-min on-source, 5-min off-source observation style. A longer observation time was chosen to maximize time on Proxima Centauri to search for flare emission: a key element which may determine the habitability of Proxima~b. Longer observations mean we are insensitive to signals lasting less than 30 minutes because we do not have intervening off-source pointings needed to discern if a hit is caused by RFI. However, we are sensitive to signals that are broadcast over a long period of time (an hour or longer) because we can see how that signal changes over time (e.g. drift rate changes). We are also more sensitive to weaker signals because we can integrate over the whole 30 minute observation to look for a persistent signal. 

Second, we used longer cadences---sets of pointings toward the target source and reference sources---than the 6-pointing default for BL. The total number of pointings in a single observation ranged from 12 to 17. This gives us the flexibility to choose a subset of cadences from a larger number of pointings per day. In our initial search, we considered 4-pointing cadences, meaning for example that if an observation consisted of 12 pointings, we looked at 9 subset cadences in that observation. 

\subsection*{Data Format}
Data are processed using the a processing pipeline run on the BLPDR. BLPDR provides data in 26 separate files, each containing a 128 MHz subband from the 0.7--4.0\,GHz band. The high-spectral-resolution products,  as used here for detection of artificial narrowband signals, have a frequency resolution of $\sim 3.81$\,Hz (i.e. $2^{25}$ channels across each 128 MHz band) and time integrations of $\sim 16.78$\,s. The final data product is stored in \verb|filterbank| format which can then be opened by the \textsc{blimpy} Python package\cite{2019JOSS....4.1554P}. The final data volume for the six day observation period is 19.5 TB: about 118$\times$ more data than were obtained for any single source in previous BL searches\cite{2020AJ....159...86P}.

 The Python/Cython package \textsc{turboSETI}\cite{Enriquez:2019} is used to search over a range of drift rates in the data\cite{2017ApJ...849..104E}. Two important parameters required by \textsc{turboSETI} are a minimum signal-to-noise ratio (S/N), and a maximum possible drift rate. The minimum S/N was set to 10, following previous work\cite{2020AJ....159...86P}. However, we tailored the drift rate range to the specific characteristics of Proxima~b's rotation and orbit. 

\subsection*{Expected Drift Rate}

The most dominant factors affecting the drift rate of a signal are the rotations and orbits of the Earth and the source body. The following equation\cite{2019ApJ...884...14S}, gives us the maximum expected Doppler drift rate ($\dot{\nu}_{\mathrm{max}}$) by accounting for planet rotation $\left ( \frac{4\pi^{2} R}{P^2} \right)$ and orbit $\left ( \frac{GM}{r} \right)$:

\begin{equation}
	\dot{\nu}_{\mathrm{max}} = \frac{\nu_{0}}{c} \left ( \frac{4\pi^{2} R_{\oplus}}{P^2_{\oplus}} + \frac{4\pi^{2} R_{\mathrm{Pb}}}{P^2_{\mathrm{Pb}}} + \frac{GM_{\odot}}{r^2_{\oplus}} + \frac{GM_{\mathrm{PC}}}{r^2_{\mathrm{Pb}}}
	\right). \label{eq:doppler}
\end{equation}

The term $v_0$ is the emitted frequency from the transmitter, $R$, $P$, $M$, and $r$ are the planetary radii, rotational periods, solar masses, and orbital radii for Earth (subscript $\oplus$) and Proxima~b (subscript Pb), respectively. Other contributions to the drift rate, such as the bodies' movement through the Milky Way, are negligible.

To limit computation time, an initial search of of $\pm 3$\,Hz\,s$^{-1}$ was performed across all frequencies. However, we expect $\dot{\nu}_{max} = 4.191$\,Hz\,s$^{-1}$  at 4000\,MHz using the parameters from Tab.~2 and Eq.~\ref{eq:doppler}. Therefore, a supplementary search over $\pm 4$\,Hz\,s$^{-1}$ from 2752 to 3648\,MHz and $\pm 5$\,Hz\,s$^{-1}$ from 3648 to 4032\,MHz was necessary to search the complete range of possible drift rates expected. Neverthless, putative transmitters orbiting Proxima~b could exhibit drift rates orders of magnitude higher\cite{Sheikh:2019}; extending to such high drift rates is computationally challenging, and we do not consider these here.

\subsection*{Finding Events}

To find candidate events, we run the hits (signals above the S/N threshold) found by \textsc{turboSETI} through a secondary pipeline which compares on-source and off-source pointings. We classify an event as any narrow-band hit which exists in an on-source pointing, but not any of the off-source observations. Typical BL SETI searches with single-dish telescopes use a cadence length of six (ABABAB, three on-source and three off-source observations); however, we use a cadence length of four (ABAB) due to our longer observation times. A shorter cadence relaxes the requirement that events are detected in all on-source observations; that is, we allow events with $\sim 1$-hr duration. Once an event is found in a cadence of four, we searched additional pointings to see if it occurs over a longer time period. Note that cadences are primarily used to as a discriminant for RFI; as the narrowband search is run separately on each pointing, longer cadences do not increase sensitivity.

\subsection*{Filtering Events}

After events which occur in a 4-pointing cadence (ABAB) are found, we generate plots which have an additional two pointings to make a 6-pointing cadence (ABABAB). A longer cadence allows visual inspection of the additional pointings for low S/N hits. For example, an intermittent signal may be present in only the on-source pointings for the four pointings that were searched, but then be present in a successive off-source pointing (see Suppl. Fig. 1). We discard events that are clearly present in the off-source observations but were not detected by the search pipeline (i.e. the event is present in successive off-source pointing, but did not meet the S/N threshold of 10, or the drift rate threshold \cite{2020AJ....159...86P}).

After an initial list of promising events with a 6-pointing cadence is found, we plot cadences of length 12. This longer cadence allows us to see the entire duration of the event and if it occurs in any off-source observations. If an event is present in any off-source observations, it is discarded as local RFI (see Suppl. Fig. 2). During this step we also discard events which share similar characteristics (drift rate, frequency, or profile) to other hits that are found in off-source observations. Finally, any candidate event that lies in the frequency range of nearby registered ground or satellite transmitters is marked as suspicious (see Suppl. Fig. 3).

Every event that passes the two rounds of visual inspection and lies within no registered transmitters is scrutinized. Extensive research is done on the frequency bands that the event lies within. We use allocation charts such as the ASRP, which contains an extensive list of the types of transmitters allocated to specific frequency bands, and the Australian Communications and Media Authority Register of Radiocommunications Licences (\url{https://web.acma.gov.au/rrl/}).

\section*{Data availability}
The data used in this work is available for download via \url{https://seti.berkeley.edu/blc1}. Correspondence and requests for other materials should be addressed to D. C. Price. 

\section*{Code availability}
Data analysis was performed using the \textsc{Blimpy}\cite{2019JOSS....4.1554P} and \textsc{turboSETI}\cite{2019ascl.soft06006E} Python packages. These codes are open-source and are available from \url{https://github.com/UCBerkeleySETI/} and the Python Package Index (\url{https://pypi.org/}). The \textsc{Blimpy} and \textsc{turboSETI} packages make use of Astropy\cite{AstropyCollaboration:2013,AstropyCollaboration:2018}, h5py\cite{collette_python_hdf5_2014}, Matplotlib\cite{hunter2007matplotlib},  Numpy\cite{Harris:2020}, and Pandas\cite{mckinney-proc-scipy-2010}.

\section*{Acknowledgements}
Breakthrough Listen is managed by the Breakthrough Initiatives, sponsored by the \href{http://breakthroughinitiatives.org}{Breakthrough Prize Foundation}. The Parkes radio telescope is part of the Australia Telescope
National Facility which is funded by the Australian Government
for operation as a National Facility managed by CSIRO. Shane Smith and Steve Croft were supported by the National Science Foundation under the Berkeley SETI Research Center REU Site Grant No. 1950897.

\section*{Author Contributions}
SS and DCP led the data analysis and are primary authors of the manuscript. SZS led in-depth analysis of BLC1. DJC, SC, DD, VG, HI, BL, CN, KIP, APVS, CIW assisted with interpretation, manuscript preparation and revision, and data analysis. ML and DM provided instrument support, managed data, and aided observations. AZ was lead observer during Parkes observations and aided manuscript preparation. JD and SPW aided manuscript preparation and provided logistical support.

\section*{Competing Interests}
The authors declare no competing interests.

\section*{Tables}

\subsection*{Table 1}

\begin{tabular}{lccc}

\hline
Band &  Hits & Events & Candidates  \\
\hline
\hline

UHF        & 869,081   & 2,566 & 1   \\ 
L           & 1,538,111 & 2,528 & 0  \\
S           & 1,952,162 & 66 & 0  \\
\hline
Parkes UWL & 4,172,702 & 5,160 & 1 \\

\hline
\end{tabular}

Total hits above our thresholds, events that occur in two ONs but no OFFs, and the final candidates that cannot be immediately attributed to RFI.\label{tab:totalstats} 

\subsection*{Table 2}

\begin{tabular}{ll}
\hline
Parameter &  Value  \\
\hline
\hline
Prox b radius $\left ( R_{\mathrm{Pb}} \right)$       & $1.07_{-0.31}^{+0.38}$ $R_{\oplus}$  \\ 
Prox b rotation $\left ( P_{\mathrm{Pb}} \right)$ & $11.18427 \pm{0.00070}$ days \\
Prox b orbit $\left ( r_{\mathrm{Pb}} \right)$        & $0.04864 \pm{0.00031}$ AU \\
Prox Cen mass $\left ( M_{\mathrm{PC}} \right)$          & $0.1221\pm{0.0022}$ $M_{\odot}$\\
\hline
\end{tabular}

\label{tab:parameters} Parameters used in the Doppler drift rate calculation from Eq. \ref{eq:doppler}. Data for Proxima Centauri and Proxima~b were taken from \cite{Su_rez_Mascare_o_2020} and \cite{2017ApJ...836L..31B}. Using the listed uncertainties, we calculate maximum drift rate error of $\dot{\nu}_{max} = $ 0.24\,mHz\,s$^{-1}$ and $\dot{\nu}_{max} = $ 1.38\,mHz\,s$^{-1}$ at 704\,MHz and 4032\,MHz respectively.



\renewcommand{\thefigure}{(Supplementary) \arabic{figure}}
\setcounter{figure}{0}

\thispagestyle{empty}


\section*{Supplementary information}

\subsection*{Filtered Events}

Here, we provide examples of events which do not pass our search criteria after three rounds of analysis. Supplementary Fig. 1 shows a candidate signal at $\sim$749.326\,MHz, which is narrowband and exhibits a variable drift rate. This candidate failed the first round of plotting tests with six cadence plots, as the signal is visible in the first OFF observation (toward PKS 1421-490). Supplementary Fig. 2 shows a candidate signal at $\sim$1000.151\,MHz, which is consistent with a non-disciplined oscillator as used in consumer digital electronics. This signal failed the second round of plotting tests (with 12 cadence plots) because the signal is clearly visible in the earlier plots; however, the signal was not detected above the S/N threshold in the earlier cadence. Supplementary Fig. 3 passed two rounds of plotting tests, but is ultimately discarded as satellite RFI from the Global Navigation Satellite System (GNSS), which is known to operate at these frequencies. 

\subsection*{Flare Searches}

During the multi-telescope, wide-band radio monitoring of space weather on Proxima Centauri campaign, three different flare events were detected (AB1, AB2, and AB3) with the Australian Square Kilometre Array Pathfinder (ASKAP). The type IV flare AB3 occurred during the observation campaign (900--1200 MHz at MBJD 58605.5652), but a few hours before Parkes observations started. To search for flares on minute to hour timescales, we averaged our high frequency resolution  Stokes-I data into $\sim$kHz resolution data products and subtracting the mean bandpass. No signs of flare activity was apparent. This is likely due to gain variations within the UWL receiver, which are difficult to discern from astrophysical variations. These gain variations can be compensated for using noise added radiometry techniques, whereby a calibrated and temperature-stabilised noise diode is injected into the UWL receiver path. However, the noise diode was not used as it corrupts the high time resolution (128 $\mu$s) data product. The high time resolution data was ultimately not analyzed for flare activity, as ASKAP and other telescopes used in the commensal campaign did not detect further flares. 

\begin{figure}[hbt]

  \centering
  \includegraphics[width=\textwidth]{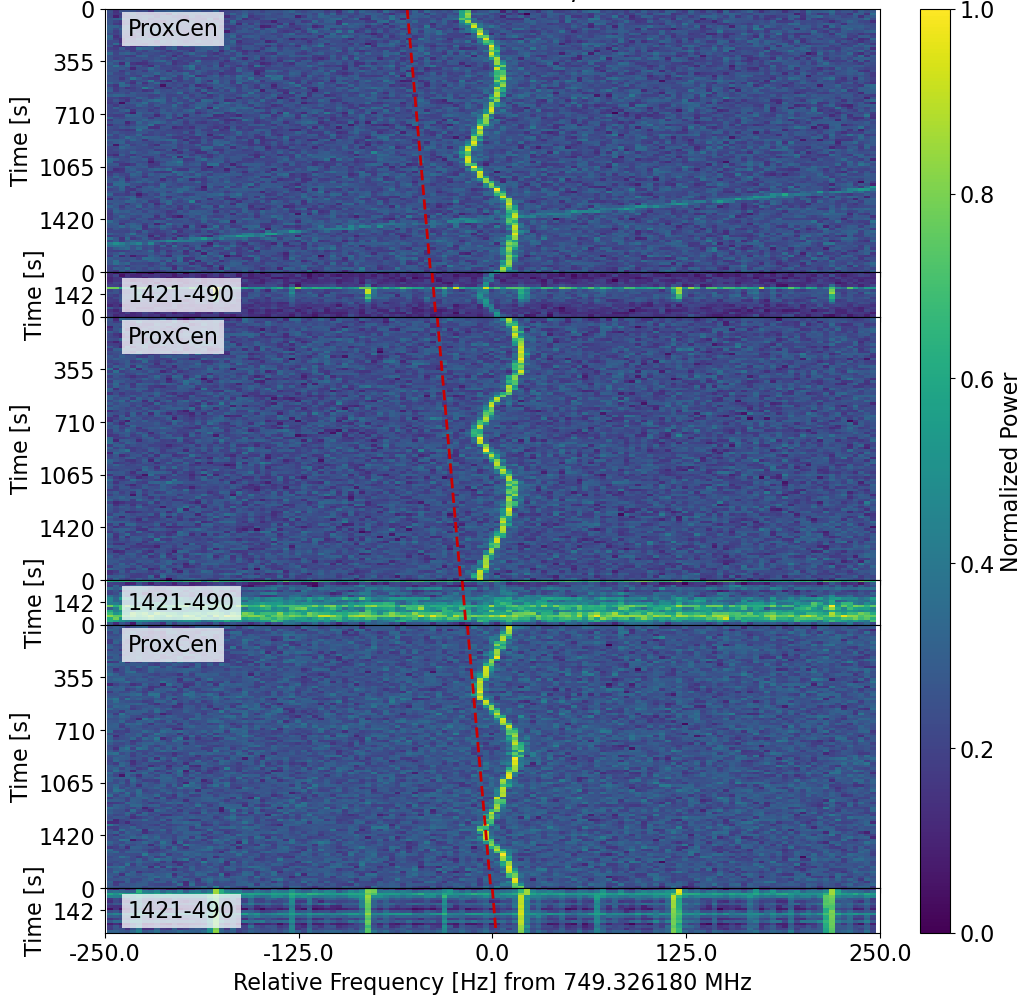}
  \caption{An example of an event that fails the first round of plotting tests (cadence length of 6). All cadence plots, including this one, contain a red dashed line which represents a left-offset estimation of the calculated drift rate. This drift estimation represents only the relative angle of the drift rate which is calculated in the first observation. The signal is clearly seen by eye in the second OFF pointing but not detected by \textsc{turboSETI} above the required signal to noise ratio.}
\end{figure}

\begin{figure}[hbt]
\centering
  \includegraphics[width=0.6\textwidth]{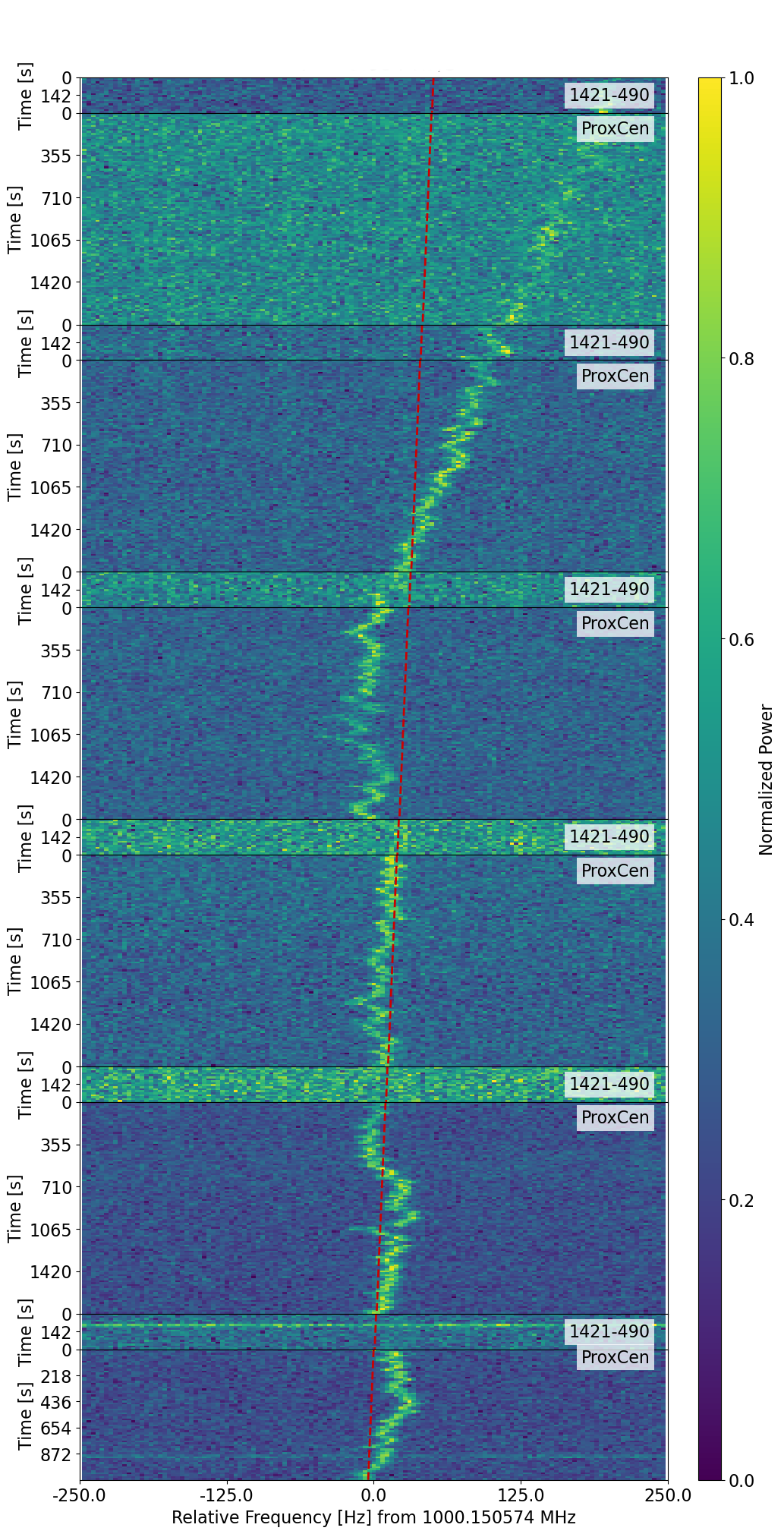}
  \caption{An event that did not pass the second round of plotting (cadence length of 12). The signal was not visible in the off-source pointings for the first round of plotting (6th--11th pointings), but the signal is clearly visible in earlier off-source pointings that were not initially plotted.}
\end{figure}

\begin{figure}[hbt]
\centering
  \includegraphics[width=0.6\textwidth]{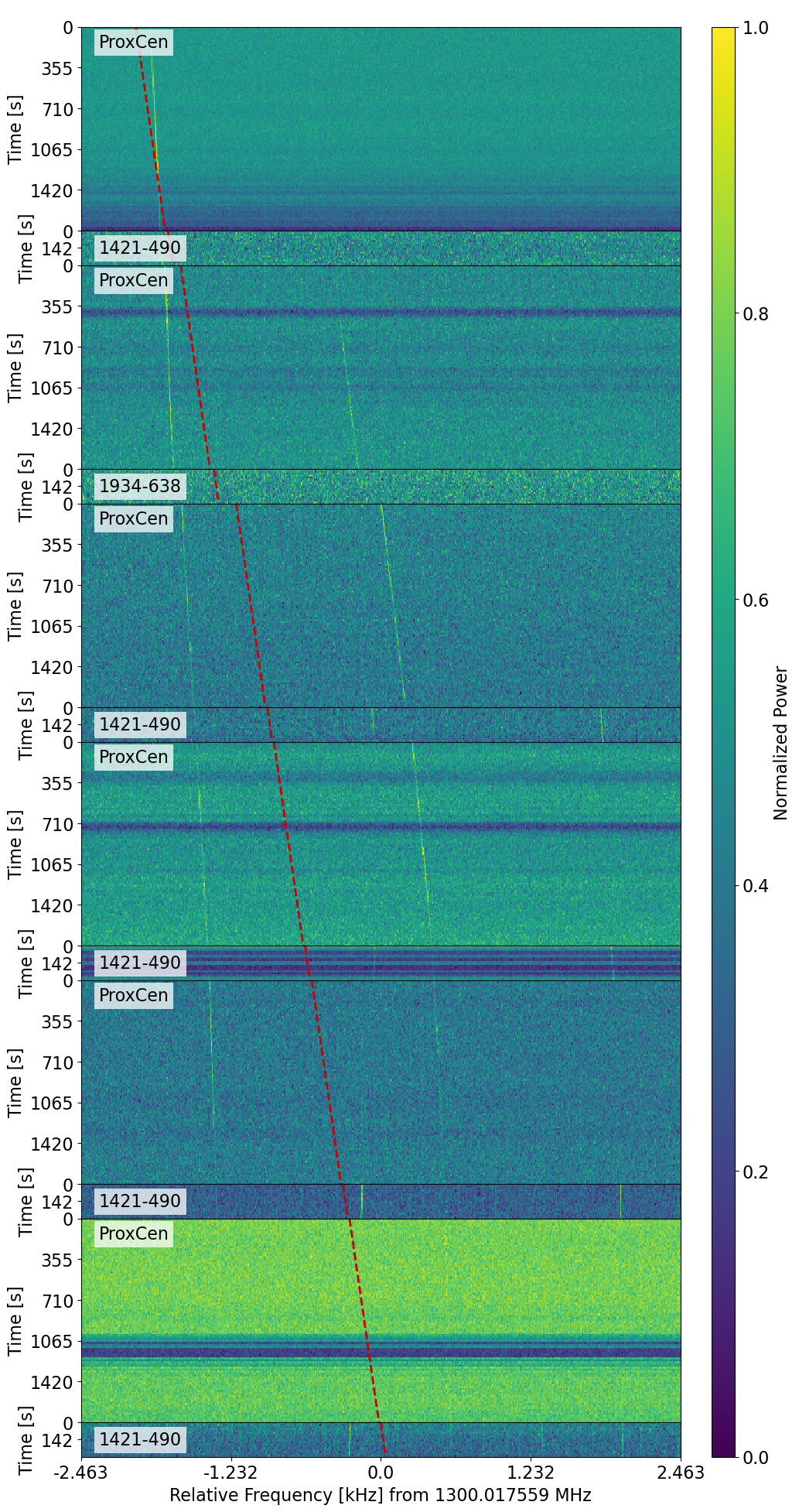}
  \caption{Example of an event that passed two rounds of plotting, but was flagged as satellite RFI. The Global Navigation Satellite Service (GNSS) emits in the $1238 - 1318$\,MHz range. In addition to the two events visible in a majority of the on-source pointings, two similar events are visible in some of the off-source pointings. This is likely due to another GNSS satellite that is close to our pointing of PKS\,1421-490.}
\end{figure}



\begin{thebibliography}{10}
\urlstyle{rm}
\expandafter\ifx\csname url\endcsname\relax
  \def\url#1{\texttt{#1}}\fi
\expandafter\ifx\csname urlprefix\endcsname\relax\def\urlprefix{URL }\fi
\expandafter\ifx\csname doiprefix\endcsname\relax\def\doiprefix{DOI: }\fi
\providecommand{\bibinfo}[2]{#2}
\providecommand{\eprint}[2][]{\url{#2}}

\bibitem{Sheikh:2021:BLC1}
\bibinfo{author}{{Sheikh}, S.} \emph{et~al.}
\newblock \bibinfo{journal}{\bibinfo{title}{{}}}.
\newblock {\emph{\JournalTitle{\nastro}}}  (\bibinfo{year}{2021}).

\bibitem{2016Natur.536..437A}
\bibinfo{author}{{Anglada-Escud{\'e}}, G.} \emph{et~al.}
\newblock \bibinfo{journal}{\bibinfo{title}{{A terrestrial planet candidate in
  a temperate orbit around Proxima Centauri}}}.
\newblock {\emph{\JournalTitle{\nat}}} \textbf{\bibinfo{volume}{536}},
  \bibinfo{pages}{437--440}, \doiprefix\url{10.1038/nature19106}
  (\bibinfo{year}{2016}).

\bibitem{2016A&A...596A.111R}
\bibinfo{author}{{Ribas}, I.} \emph{et~al.}
\newblock \bibinfo{journal}{\bibinfo{title}{{The habitability of Proxima
  Centauri b. I. Irradiation, rotation and volatile inventory from formation to
  the present}}}.
\newblock {\emph{\JournalTitle{\aap}}} \textbf{\bibinfo{volume}{596}},
  \bibinfo{pages}{A111}, \doiprefix\url{10.1051/0004-6361/201629576}
  (\bibinfo{year}{2016}).

\bibitem{2016A&A...596A.112T}
\bibinfo{author}{{Turbet}, M.} \emph{et~al.}
\newblock \bibinfo{journal}{\bibinfo{title}{{The habitability of Proxima
  Centauri b. II. Possible climates and observability}}}.
\newblock {\emph{\JournalTitle{\aap}}} \textbf{\bibinfo{volume}{596}},
  \bibinfo{pages}{A112}, \doiprefix\url{10.1051/0004-6361/201629577}
  (\bibinfo{year}{2016}).


\bibitem{Alvarado-Gomez:2020}
\bibinfo{author}{{Alvarado-G{\'o}mez}, J.~D.} \emph{et~al.}
\newblock \bibinfo{journal}{\bibinfo{title}{{An Earth-like Stellar Wind
  Environment for Proxima Centauri c}}}.
\newblock {\emph{\JournalTitle{\apjl}}} \textbf{\bibinfo{volume}{902}},
  \bibinfo{pages}{L9}, \doiprefix\url{10.3847/2041-8213/abb885}
  (\bibinfo{year}{2020}).


\bibitem{2018ApJ...860L..30H}
\bibinfo{author}{{Howard}, W.~S.} \emph{et~al.}
\newblock \bibinfo{journal}{\bibinfo{title}{{The First Naked-eye Superflare
  Detected from Proxima Centauri}}}.
\newblock {\emph{\JournalTitle{\apjl}}} \textbf{\bibinfo{volume}{860}},
  \bibinfo{pages}{L30}, \doiprefix\url{10.3847/2041-8213/aacaf3}
  (\bibinfo{year}{2018}).


\bibitem{Zic:2020}
\bibinfo{author}{{Zic}, A.} \emph{et~al.}
\newblock \bibinfo{journal}{\bibinfo{title}{{A Flare-type IV Burst Event from
  Proxima Centauri and Implications for Space Weather}}}.
\newblock {\emph{\JournalTitle{\apj}}} \textbf{\bibinfo{volume}{905}},
  \bibinfo{pages}{23}, \doiprefix\url{10.3847/1538-4357/abca90}
  (\bibinfo{year}{2020}).


\bibitem{MacGregor:2021}
\bibinfo{author}{{MacGregor}, M.~A.} \emph{et~al.}
\newblock \bibinfo{journal}{\bibinfo{title}{{Discovery of an Extremely Short
  Duration Flare from Proxima Centauri Using Millimeter through Far-ultraviolet
  Observations}}}.
\newblock {\emph{\JournalTitle{\apjl}}} \textbf{\bibinfo{volume}{911}},
  \bibinfo{pages}{L25}, \doiprefix\url{10.3847/2041-8213/abf14c}
  (\bibinfo{year}{2021}).
  
\bibitem{Tarter:2007}
\bibinfo{author}{{Tarter}, J.~C.} \emph{et~al.}
\newblock \bibinfo{journal}{\bibinfo{title}{{A Reappraisal of The Habitability
  of Planets around M Dwarf Stars}}}.
\newblock {\emph{\JournalTitle{Astrobiology}}} \textbf{\bibinfo{volume}{7}},
  \bibinfo{pages}{30--65}, \doiprefix\url{10.1089/ast.2006.0124}
  (\bibinfo{year}{2007}).

\bibitem{Parkin:2018}
\bibinfo{author}{{Parkin}, K. L.~G.}
\newblock \bibinfo{journal}{\bibinfo{title}{{The Breakthrough Starshot system
  model}}}.
\newblock {\emph{\JournalTitle{Acta Astronautica}}}
  \textbf{\bibinfo{volume}{152}}, \bibinfo{pages}{370--384},
  \doiprefix\url{10.1016/j.actaastro.2018.08.035} (\bibinfo{year}{2018}).

\bibitem{Backus:1995}
\bibinfo{author}{{Backus}, P.~R.}
\newblock \bibinfo{title}{{Project PHOENIX SETI Observations at Parkes}}.
\newblock In \emph{\bibinfo{booktitle}{American Astronomical Society Meeting
  Abstracts}}, vol. \bibinfo{volume}{187} of \emph{\bibinfo{series}{American
  Astronomical Society Meeting Abstracts}}, \bibinfo{pages}{41.01}
  (\bibinfo{year}{1995}).

\bibitem{Tarter:1996}
\bibinfo{author}{{Tarter}, J.~C.}
\newblock \bibinfo{title}{{Project Phoenix: the Australian deployment}}.
\newblock In \bibinfo{editor}{{Kingsley}, S.~A.} \&
  \bibinfo{editor}{{Lemarchand}, G.~A.} (eds.) \emph{\bibinfo{booktitle}{The
  Search for Extraterrestrial Intelligence (SETI) in the Optical Spectrum II}},
  vol. \bibinfo{volume}{2704} of \emph{\bibinfo{series}{Society of
  Photo-Optical Instrumentation Engineers (SPIE) Conference Series}},
  \bibinfo{pages}{24--34}, \doiprefix\url{10.1117/12.243444}
  (\bibinfo{year}{1996}).

\bibitem{Blair:1992}
\bibinfo{author}{{Blair}, D.~G.} \emph{et~al.}
\newblock \bibinfo{journal}{\bibinfo{title}{{A narrow-band search for
  extraterrestrial intelligence (SETI) using the interstellar contact channel
  hypothesis.}}}
\newblock {\emph{\JournalTitle{\mnras}}} \textbf{\bibinfo{volume}{257}},
  \bibinfo{pages}{105--109}, \doiprefix\url{10.1093/mnras/257.1.105}
  (\bibinfo{year}{1992}).

\bibitem{Marcy:2021}
\bibinfo{author}{{Marcy}, G.~W.}
\newblock \bibinfo{journal}{\bibinfo{title}{{A Search for Optical Laser
  Emission from Proxima Centauri}}}.
\newblock {\emph{\JournalTitle{arXiv e-prints}}}
  \bibinfo{pages}{arXiv:2102.01910} (\bibinfo{year}{2021}).


\bibitem{2017AcAau.139...98W}
\bibinfo{author}{{Worden}, S.~P.} \emph{et~al.}
\newblock \bibinfo{journal}{\bibinfo{title}{{Breakthrough Listen - A new search
  for life in the universe}}}.
\newblock {\emph{\JournalTitle{Acta Astronautica}}}
  \textbf{\bibinfo{volume}{139}}, \bibinfo{pages}{98--101},
  \doiprefix\url{10.1016/j.actaastro.2017.06.008} (\bibinfo{year}{2017}).

\bibitem{2017PASP..129e4501I}
\bibinfo{author}{{Isaacson}, H.} \emph{et~al.}
\newblock \bibinfo{journal}{\bibinfo{title}{{The Breakthrough Listen Search for
  Intelligent Life: Target Selection of Nearby Stars and Galaxies}}}.
\newblock {\emph{\JournalTitle{\pasp}}} \textbf{\bibinfo{volume}{129}},
  \bibinfo{pages}{054501}, \doiprefix\url{10.1088/1538-3873/aa5800}
  (\bibinfo{year}{2017}).

\bibitem{Enriquez:2017}
\bibinfo{author}{{Enriquez}, J.~E.} \emph{et~al.}
\newblock \bibinfo{journal}{\bibinfo{title}{{The Breakthrough Listen Search for
  Intelligent Life: 1.1-1.9 GHz Observations of 692 Nearby Stars}}}.
\newblock {\emph{\JournalTitle{\apj}}} \textbf{\bibinfo{volume}{849}},
  \bibinfo{pages}{104}, \doiprefix\url{10.3847/1538-4357/aa8d1b}
  (\bibinfo{year}{2017}).

\bibitem{2020AJ....159...86P}
\bibinfo{author}{{Price}, D.~C.} \emph{et~al.}
\newblock \bibinfo{journal}{\bibinfo{title}{{The Breakthrough Listen Search for
  Intelligent Life: Observations of 1327 Nearby Stars Over 1.10--3.45 GHz}}}.
\newblock {\emph{\JournalTitle{\aj}}} \textbf{\bibinfo{volume}{159}},
  \bibinfo{pages}{86}, \doiprefix\url{10.3847/1538-3881/ab65f1}
  (\bibinfo{year}{2020}).

\bibitem{Sheikh:2020}
\bibinfo{author}{{Sheikh}, S.~Z.} \emph{et~al.}
\newblock \bibinfo{journal}{\bibinfo{title}{{The Breakthrough Listen Search for
  Intelligent Life: A 3.95-8.00 GHz Search for Radio Technosignatures in the
  Restricted Earth Transit Zone}}}.
\newblock {\emph{\JournalTitle{\aj}}} \textbf{\bibinfo{volume}{160}},
  \bibinfo{pages}{29}, \doiprefix\url{10.3847/1538-3881/ab9361}
  (\bibinfo{year}{2020}).

\bibitem{Traas:2021}
\bibinfo{author}{{Traas}, R.} \emph{et~al.}
\newblock \bibinfo{journal}{\bibinfo{title}{{The Breakthrough Listen Search for
  Intelligent Life: Searching for Technosignatures in Observations of TESS
  Targets of Interest}}}.
\newblock {\emph{\JournalTitle{\aj}}} \textbf{\bibinfo{volume}{161}},
  \bibinfo{pages}{286}, \doiprefix\url{10.3847/1538-3881/abf649}
  (\bibinfo{year}{2021}).

\bibitem{Gajjar:2021}
\bibinfo{author}{{Gajjar}, V.} \emph{et~al.}
\newblock \bibinfo{journal}{\bibinfo{title}{{The Breakthrough Listen Search For
  Intelligent Life Near the Galactic Center. I.}}}
\newblock {\emph{\JournalTitle{\aj}}} \textbf{\bibinfo{volume}{162}},
  \bibinfo{pages}{33}, \doiprefix\url{10.3847/1538-3881/abfd36}
  (\bibinfo{year}{2021}).

\bibitem{2006JNav...59..293Z}
\bibinfo{author}{{Zhang}, J.}, \bibinfo{author}{{Zhang}, K.},
  \bibinfo{author}{{Grenfell}, R.} \& \bibinfo{author}{{Deakin}, R.}
\newblock \bibinfo{journal}{\bibinfo{title}{{GPS Satellite Velocity and
  Acceleration Determination using the Broadcast Ephemeris}}}.
\newblock {\emph{\JournalTitle{Journal of Navigation}}}
  \textbf{\bibinfo{volume}{59}}, \bibinfo{pages}{293--305},
  \doiprefix\url{10.1017/S0373463306003638} (\bibinfo{year}{2006}).

\bibitem{Forgan:2019}
\bibinfo{author}{{Forgan}, D.} \emph{et~al.}
\newblock \bibinfo{journal}{\bibinfo{title}{{Rio 2.0: revising the Rio scale
  for SETI detections}}}.
\newblock {\emph{\JournalTitle{International Journal of Astrobiology}}}
  \textbf{\bibinfo{volume}{18}}, \bibinfo{pages}{336--344},
  \doiprefix\url{10.1017/S1473550418000162} (\bibinfo{year}{2019}).

\bibitem{2017ApJ...849..104E}
\bibinfo{author}{{Enriquez}, J.~E.} \emph{et~al.}
\newblock \bibinfo{journal}{\bibinfo{title}{{The Breakthrough Listen Search for
  Intelligent Life: 1.1-1.9 GHz Observations of 692 Nearby Stars}}}.
\newblock {\emph{\JournalTitle{\apj}}} \textbf{\bibinfo{volume}{849}},
  \bibinfo{pages}{104}, \doiprefix\url{10.3847/1538-4357/aa8d1b}
  (\bibinfo{year}{2017}).

\bibitem{Wlodarczyk-Sroka:2020}
\bibinfo{author}{{Wlodarczyk-Sroka}, B.~S.}, \bibinfo{author}{{Garrett}, M.~A.}
  \& \bibinfo{author}{{Siemion}, A.~P.~V.}
\newblock \bibinfo{journal}{\bibinfo{title}{{Extending the Breakthrough Listen
  nearby star survey to other stellar objects in the field}}}.
\newblock {\emph{\JournalTitle{\mnras}}} \textbf{\bibinfo{volume}{498}},
  \bibinfo{pages}{5720--5729}, \doiprefix\url{10.1093/mnras/staa2672}
  (\bibinfo{year}{2020}).

\bibitem{2020PASA...37...12H}
\bibinfo{author}{{Hobbs}, G.} \emph{et~al.}
\newblock \bibinfo{journal}{\bibinfo{title}{{An ultra-wide bandwidth (704 to 4
  032 MHz) receiver for the Parkes radio telescope}}}.
\newblock {\emph{\JournalTitle{\pasa}}} \textbf{\bibinfo{volume}{37}},
  \bibinfo{pages}{e012}, \doiprefix\url{10.1017/pasa.2020.2}
  (\bibinfo{year}{2020}).

\bibitem{Price:2018}
\bibinfo{author}{{Price}, D.~C.} \emph{et~al.}
\newblock \bibinfo{journal}{\bibinfo{title}{{The Breakthrough Listen search for
  intelligent life: Wide-bandwidth digital instrumentation for the CSIRO Parkes
  64-m telescope}}}.
\newblock {\emph{\JournalTitle{\pasa}}} \textbf{\bibinfo{volume}{35}},
  \bibinfo{pages}{41}, \doiprefix\url{10.1017/pasa.2018.36}
  (\bibinfo{year}{2018}).

\bibitem{Price:2021}
\bibinfo{author}{{Price}, D.~C.} \emph{et~al.}
\newblock \bibinfo{journal}{\bibinfo{title}{{Expanded Capability of the
  Breakthrough Listen Parkes Data Recorder for Observations with the UWL
  Receiver}}}.
\newblock {\emph{\JournalTitle{Research Notes of the American Astronomical
  Society}}} \textbf{\bibinfo{volume}{5}}, \bibinfo{pages}{114},
  \doiprefix\url{10.3847/2515-5172/ac00c1} (\bibinfo{year}{2021}).

\bibitem{2019JOSS....4.1554P}
\bibinfo{author}{{Price}, D.}, \bibinfo{author}{{Enriquez}, J.},
  \bibinfo{author}{{Chen}, Y.} \& \bibinfo{author}{{Siebert}, M.}
\newblock \bibinfo{journal}{\bibinfo{title}{{Blimpy: Breakthrough Listen I/O
  Methods for Python}}}.
\newblock {\emph{\JournalTitle{The Journal of Open Source Software}}}
  \textbf{\bibinfo{volume}{4}}, \bibinfo{pages}{1554},
  \doiprefix\url{10.21105/joss.01554} (\bibinfo{year}{2019}).

\bibitem{Enriquez:2019}
\bibinfo{author}{{Enriquez}, E.} \& \bibinfo{author}{{Price}, D.}
\newblock \bibinfo{journal}{\bibinfo{title}{{turboSETI: Python-based SETI
  search algorithm}}}.
\newblock {\emph{\JournalTitle{Astrophysics Source Code Library}}}
  \bibinfo{pages}{ascl:1906.006} (\bibinfo{year}{2019}).

\bibitem{2019ApJ...884...14S}
\bibinfo{author}{{Sheikh}, S.~Z.}, \bibinfo{author}{{Wright}, J.~T.},
  \bibinfo{author}{{Siemion}, A.} \& \bibinfo{author}{{Enriquez}, J.~E.}
\newblock \bibinfo{journal}{\bibinfo{title}{{Choosing a Maximum Drift Rate in a
  SETI Search: Astrophysical Considerations}}}.
\newblock {\emph{\JournalTitle{\apj}}} \textbf{\bibinfo{volume}{884}},
  \bibinfo{pages}{14}, \doiprefix\url{10.3847/1538-4357/ab3fa8}
  (\bibinfo{year}{2019}).

\bibitem{Sheikh:2019}
\bibinfo{author}{{Sheikh}, S.~Z.}, \bibinfo{author}{{Wright}, J.~T.},
  \bibinfo{author}{{Siemion}, A.} \& \bibinfo{author}{{Enriquez}, J.~E.}
\newblock \bibinfo{journal}{\bibinfo{title}{{Choosing a Maximum Drift Rate in a
  SETI Search: Astrophysical Considerations}}}.
\newblock {\emph{\JournalTitle{\apj}}} \textbf{\bibinfo{volume}{884}},
  \bibinfo{pages}{14}, \doiprefix\url{10.3847/1538-4357/ab3fa8}
  (\bibinfo{year}{2019}).

\bibitem{2019ascl.soft06006E}
\bibinfo{author}{{Enriquez}, E.} \& \bibinfo{author}{{Price}, D.}
\newblock \bibinfo{title}{{turboSETI: Python-based SETI search algorithm}}
  (\bibinfo{year}{2019}).
\newblock \eprint{ascl:1906.006}.

\bibitem{AstropyCollaboration:2013}
\bibinfo{author}{{Astropy Collaboration}} \emph{et~al.}
\newblock \bibinfo{journal}{\bibinfo{title}{{Astropy: A community Python
  package for astronomy}}}.
\newblock {\emph{\JournalTitle{\aap}}} \textbf{\bibinfo{volume}{558}},
  \bibinfo{pages}{A33}, \doiprefix\url{10.1051/0004-6361/201322068}
  (\bibinfo{year}{2013}).

\bibitem{AstropyCollaboration:2018}
\bibinfo{author}{{Astropy Collaboration}} \emph{et~al.}
\newblock \bibinfo{journal}{\bibinfo{title}{{The Astropy Project: Building an
  Open-science Project and Status of the v2.0 Core Package}}}.
\newblock {\emph{\JournalTitle{\aj}}} \textbf{\bibinfo{volume}{156}},
  \bibinfo{pages}{123}, \doiprefix\url{10.3847/1538-3881/aabc4f}
  (\bibinfo{year}{2018}).

\bibitem{collette_python_hdf5_2014}
\bibinfo{author}{Collette, A.}
\newblock \emph{\bibinfo{title}{Python and HDF5}}
  (\bibinfo{publisher}{O'Reilly}, \bibinfo{year}{2013}).

\bibitem{hunter2007matplotlib}
\bibinfo{author}{Hunter, J.~D.}
\newblock \bibinfo{journal}{\bibinfo{title}{Matplotlib: A 2d graphics
  environment}}.
\newblock {\emph{\JournalTitle{Computing in science \& engineering}}}
  \textbf{\bibinfo{volume}{9}}, \bibinfo{pages}{90--95} (\bibinfo{year}{2007}).

\bibitem{Harris:2020}
\bibinfo{author}{{Harris}, C.~R.} \emph{et~al.}
\newblock \bibinfo{journal}{\bibinfo{title}{{Array programming with NumPy}}}.
\newblock {\emph{\JournalTitle{\nat}}} \textbf{\bibinfo{volume}{585}},
  \bibinfo{pages}{357--362}, \doiprefix\url{10.1038/s41586-020-2649-2}
  (\bibinfo{year}{2020}).

\bibitem{mckinney-proc-scipy-2010}
\bibinfo{author}{{W}es {M}c{K}inney}.
\newblock \bibinfo{title}{{D}ata {S}tructures for {S}tatistical {C}omputing in
  {P}ython}.
\newblock In \bibinfo{editor}{{S}t\'efan van~der {W}alt} \&
  \bibinfo{editor}{{J}arrod {M}illman} (eds.)
  \emph{\bibinfo{booktitle}{{P}roceedings of the 9th {P}ython in {S}cience
  {C}onference}}, \bibinfo{pages}{56 -- 61},
  \doiprefix\url{10.25080/Majora-92bf1922-00a} (\bibinfo{year}{2010}).

\bibitem{Su_rez_Mascare_o_2020}
\bibinfo{author}{Suárez~Mascareño, A.} \emph{et~al.}
\newblock \bibinfo{journal}{\bibinfo{title}{Revisiting proxima with espresso}}.
\newblock {\emph{\JournalTitle{\aap}}} \textbf{\bibinfo{volume}{639}},
  \bibinfo{pages}{A77}, \doiprefix\url{10.1051/0004-6361/202037745}
  (\bibinfo{year}{2020}).

\bibitem{2017ApJ...836L..31B}
\bibinfo{author}{{Bixel}, A.} \& \bibinfo{author}{{Apai}, D.}
\newblock \bibinfo{journal}{\bibinfo{title}{{Probabilistic Constraints on the
  Mass and Composition of Proxima b}}}.
\newblock {\emph{\JournalTitle{\apjl}}} \textbf{\bibinfo{volume}{836}},
  \bibinfo{pages}{L31}, \doiprefix\url{10.3847/2041-8213/aa5f51}
  (\bibinfo{year}{2017}).

\end{thebibliography}

\end{document}